\documentclass[10pt]{article}
\usepackage[T1]{fontenc}
\usepackage[latin1]{inputenc}
\usepackage{graphics}
\usepackage{setspace}
\makeatletter

\newcommand{\LyX}{L\kern-.1667em\lower.25em\hbox{Y}\kern-.125emX\spacefactor1000}

\newcommand{\lyxaddress}[1]{
  \par {\raggedright #1 
  \vspace{1.4em}
  \noindent\par}
}

\makeatother

\begin{document}

\title{Unsupervised learning and adaptation in a model of adult neurogenesis}

\author{Guillermo A. Cecchi\protect\( ^{1}\protect \), Leopoldo T. Petreanu\protect\( ^{2}\protect \),
\\
Arturo Alvarez-Buylla\protect\( ^{2}\protect \) and Marcelo O. Magnasco\protect\( ^{1}\protect \) }

\maketitle

\lyxaddress{Laboratories of \protect\( ^{1}\protect \)Mathematical Physics and \protect\( ^{2}\protect \)
Neurogenesis, The Rockefeller University, 1230 York Avenue, New York NY10021,
U.S.A.}

\begin{abstract}
Adult neurogenesis has long been documented in the vertebrate brain, and recently
even in humans. Although it has been conjectured for many years that its functional
role is related to the renewing of memories, no clear mechanism as to how this
can be achieved has been proposed. We present a scheme in which incorporation
of new neurons proceeds at a constant rate, while their survival is activity-dependent
and thus contingent upon new neurons establishing suitable connections. We show
that a simple mathematical model following these rules organizes its activity
so as to maximize the difference between its responses, and can adapt to changing
environmental conditions in unsupervised fashion.
\end{abstract}

\section{Introduction:}

The function of the mature brain is not only to be constantly ready to solve
particular tasks, but also to learn new ones, adapt to changes in environmental
conditions, and optimize its computational resources accordingly. While synaptic
change is regarded as the most common form of modification underlying plasticity
(notably in neural network models), it is by no means the only source of functional
change in the brain. One alternate source is adult neurogenesis, i.e. the incorporation
of new neurons into neural structures that have already passed through development.

It was long thought that neuronal production ended soon after birth, fueling
a view of the adult brain as a finalized structure (as far as new neurons are
concerned), with further plasticity concentrated on synapses. It took some time
to realize and demonstrate that neuronal incorporation can, to different degrees,
continue during the entire life of an animal. De-novo neuronal incorporation
continues in the adult mammalian brain in the dentate gyrus of the hippocampus,
the olfactory bulb and the association cortex \cite{ALTMAN,LOIS,ARTURO95,GOULD1,ERIKSSON,GOULD3}.
Adult neurogenesis also occurs extensively in birds and other vertebrates \cite{FERNANDO93,GOLDMAN}. 

Based on the experimental evidence, it has been conjectured that adult neurogenesis
could be related to brain plasticity and learning \cite{ARTURO90,KIRN,BARNEA,GOULD2,KEMPERMANN}.
However, the functional mechanism of neuronal replacement remains elusive, and
we do not understand the degree to which it may supersede, compete with or complement
better understood plastic mechanisms. The observation of seasonal changes of
neuronal incorporation and death in canaries, as they renew their yearly repertoire
of songs, has led to the idea that neurons are not indefinitely plastic but
that new neurons are needed for new memories. Despite the considerable interest
and importance of the phenomenon, no clear mechanism has been described through
which this memory renewal process would be implemented.

We describe here a simple model in which adult neurogenesis underlies unsupervised
learning and adaptation in sensory processing. We chose the olfactory bulb (OB)
as the neuroanatomical locus of the model, for adult neurogenesis has been extensively
documented in this system, and the neuronal processing tasks are presumably
better circumscribed than in other sensory systems \cite{MORI}. The olfactory
bulb is the first realy of the olfactory information. Olfactory receptor neurons
express a single olfactory receptor and project to two glomeruli of each bulb
in a topographically precise manner \cite{VASSAR,MOMBA}. Each receptor may
be activated by more than one odor and each odor may activate more than one
receptor \cite{MALNIC,HILDEB}. Mitral and tufted cells (M/T), the output neurons
in the bulb, receive their synaptic input from the olfactory receptors within
single glomeruli. Odor information seems to be encoded in the spatial as well
as the temporal responses of M/T cells. Olfactory discrimination is then dependent
on the combinatorial pattern of activation of multiple M/T cells. Given the
promiscuous nature of olfactory receptors, the discrimination of odors is importantly
dependent on the tuning of the combinatorial response of M/T cells. 

The activity of M/T cells is regulated by inhibitory granular cells. Granular
cells are unpolarized neurons that establish inhibitory dendro-dendritic connections
with the M/T cells long secondary dendrites\cite{WOOLF}, and are indeed thought
to play an important role in lateral inhibition and in the tuning of M/T cell
responses to different odors \cite{YOKOI}. Each granule cell connects to several
mitral cells that can be far away fron each other given the extensive lenght
of their secondary dendrites. Experimental results indicate that only inhibitory
neurons are replaced in the OB. Whereas M/T cells are born before birth, the
large majority of granular neurons continue to be born postnatally and into
adult life \cite{ROSSELLI}. 

Based on the known anatomy and physiology of the OB, we modelled neurogenesis
in this structure. It is important to remark that we are not presenting an exhaustive
model of olfaction \cite{HOPFIELD}, which would require a delicate understanding
of the temporal dynamics of encoding \cite{LAURENT}, but rather using it to
demonstrate the computational potential of neurogenesis for information processing.

\section*{Model}

We base our model on two fundamental observations. First, neuronal production
in the sub-ventricular zone (SVZ) and migration toward the OB continue even
in the absence of the bulb \cite{LOIS}; this suggests that cell birth and migration
are not dependent on olfaction or activity within the bulb \cite{KIRSCH}. Second,
most newly incorporated neurons die promptly, suggesting that those that do
survive have achieved some elusive goal \cite{LEOPOLDO,KIRN2}. Neuronal activity
is known to be necessary for survival during brain development but not for the
initial synapse formation\cite{KATZ,KO}. Moreover, activity-dependent mechanisms
also seem to play an important role in the recruitment or survival of postnatally
generated OB interneurons \cite{COROTTO,FRAZIER,CUMMINGS}. These findings suggest
that the survival of the newly-generated neurons is regulated through activity. 

Thus we postulate the basic mechanism of the model: neuronal incorporation proceeds
at a constant unregulated rate, whereas neuronal survival is modulated through
activity-dependent apoptosis. New neurons arrive and establish random connections;
those with useless connections fail to participate in relevant processing and
soon die off, but neurons that establish useful connections in the context of
the network can survive and are therefore incorporated into the final structure.
Exactly which feature of neuronal activity ensures survival can then dictate
the final function of the network. 

\begin{figure}[htbp]

\vspace{0.3cm}
{\centering \resizebox*{0.5\columnwidth}{!}{\includegraphics{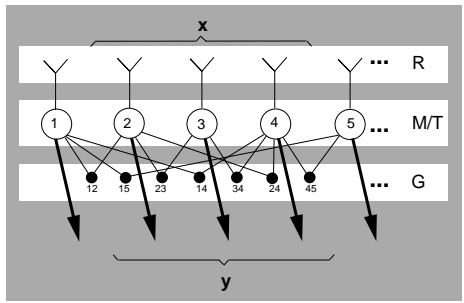}} \par}
\vspace{0.3cm}

\caption{{\footnotesize Simplified anatomy of the Olfactory Bulb used in the model. The vector \( {\bf x} \) represents the activity of the receptor layer R, and \( {\bf y} \) the activity of the mitral cell layer M/T. Granular inhibitory cells G establish pairwise connections with the mitral cells.} }\end{figure}

The simplest model compatible with the experimental evidence consists of a layer
of glomeruli receiving one-to-one projections from an input or receptor layer
and local lateral inhibition due to granular interneurons. The output of the
system then equals the input minus the effect of inhibitory connections between
the outputs:

\begin{equation}
\label{eq2}
{\bf y}^{(k)}={\bf x}^{(k)}-G{\bf y}^{(k)}
\end{equation}

\noindent where \( {\bf y}^{(k)} \) is the activity of the output layer, \( {\bf x}^{(k)} \)
is the activity of the receptor layer when presented with input \( k \), \( G \)
is a matrix giving lateral inhibition, and (\( y_{i} \) is the activity of
glomerulus number \( i \), Fig. 1). Solving for \( {\bf y} \) we obtain 
\begin{equation}
\label{eq3}
{\bf y}^{(k)}=(1+G)^{-1}{\bf x}^{(k)}
\end{equation}
We assume that each granular cell has connections to only two projection neurons
\cite{COMBO}; therefore, \( G_{ij} \) is the density or number of granular
neurons connecting M/T cells (or glomeruli) \( i \) and \( j \), so that \( G_{ij}=G_{ji} \)
\( \forall i,j \) (connections are symmetric) and \( G_{ii}=0 \) \( \forall i \),
(there is no self-inhibition). We also assume that the response of each M/T
cell is normalized by the total number of synaptic connections \cite{CHECHIK}
and that neuronal responses remain in the linear regime; positive or negative
activity should be interpreted in the context of a background level. Eq. 3 can
then be rewritten as 
\[
y^{(k)}_{i}=\frac{x^{(k)}_{i}-\sum ^{j}G_{ij}(y^{(k)}_{i}+y^{(k)}_{j})}{1+\sum _{j}G_{ij}}\]
which is explicitly normalized by total synaptic weight. We finally assume for
convenience that the output layer is normalized, \( \parallel {\bf y}^{(k)}\parallel =1\, \forall k \).

We assume that only interneurons are replaced, arriving at a constant rate with
randomly distributed connectivity. The regulation of the survival rate of the
interneurons is implemented as a function of their activity, which indicates
how well they are connected to active output cells, so that active cells survive
and inactive cells do not. Our key assumption is that the survival of the granular
cells depends on the product of the inputs, so that granular cells connecting
glomeruli \( i \) and \( j \) will survive if the value of \( y_{i}y_{j} \),
averaged over the time of presentation of the entire set of stimuli, is greater
than or equal to zero; this amounts to a somatic coincidence rule. Thus, when
\( \langle y_{i}y_{j}\rangle \gg 0 \) all newly arrived neurons survive, and
so \( \Delta G_{ij} \) has its full value; while \( \langle y_{i}y_{j}\rangle <0 \)
results in massive death and \( \Delta G_{ij}<0 \). This leads to a simplified
update equation given by
\begin{equation}
\label{eq4}
\Delta G_{ij}\sim \langle y_{i}y_{j}\rangle ,\qquad (i\neq j)
\end{equation}
where \( \langle \rangle  \) denotes the average over the ensemble of inputs
\( k=1...N \). A non-linearity is introduced through the condition that \( G \)
be positive definite, \( G_{ij}>0\, i\neq j \), i.e. the population of granular
cells cannot be negative.

\section*{Results}

Based on the previous considerations, we implemented a simulation of the process
of \emph{training} the network with the activity based replacement of the inhibitory
interneurons. The left panel of Fig. 2 shows a set of 10 odors represented as
different mixtures of receptor bindings. The right panel shows the result of
iterating a cycle of presentation of the set of odors followed by neuronal replacement
on a model network with 10 M/T cells. The result of the update rule specified
by Eq.\ref{eq4} is to \emph{orthogonalize} the output, which can be seen by
analyzing the fixed point of the system: \( \Delta {\bf y}=0\, \Rightarrow \, \Delta G_{ij}=\langle y_{i}y_{j}\rangle =0\, \forall i,j \)
and given the normalization of \( {\bf y}_{i} \), \( \langle y_{i}y_{j}\rangle \sim \bf I \),
where \( \bf I \) is the identity matrix. The simulation displays a very sharp
orthogonalization of the output vectors, which can be further appreciated in
the evolution of the response to a single odor (Fig. 3). The orthogonalization
of the output maximizes, to linear order, the separation between output vectors
and is the simplest way of achieving discrimination. An orthogonal set of neural
responses is robust to the presence of noise because the representatives are
maximally distant from each other \cite{SHANNON}. In this case, an orthogonal
set of responses from the output of the M/T cell layers corresponds to the maximal
difference in the the spatial response in the layer for each odor. So it is
possible to implement a simple algorithm based on neurogenesis and neuronal
replacement that achieves a complex mapping of input stimuli in neural space.

To quantify the orthogonalization process, we computed the determinant of the
\( 10\times 10 \) matrix of odors and M/T cells (see Methods). The evolution
of the determinant and the total number of granular cells incorporated are depicted
in Fig. 4, together with the rate of incorporation, as a function of the iteration
number.

\begin{figure}[htbp]

\vspace{0.3cm}
{\centering \resizebox*{0.45\textwidth}{!}{\includegraphics{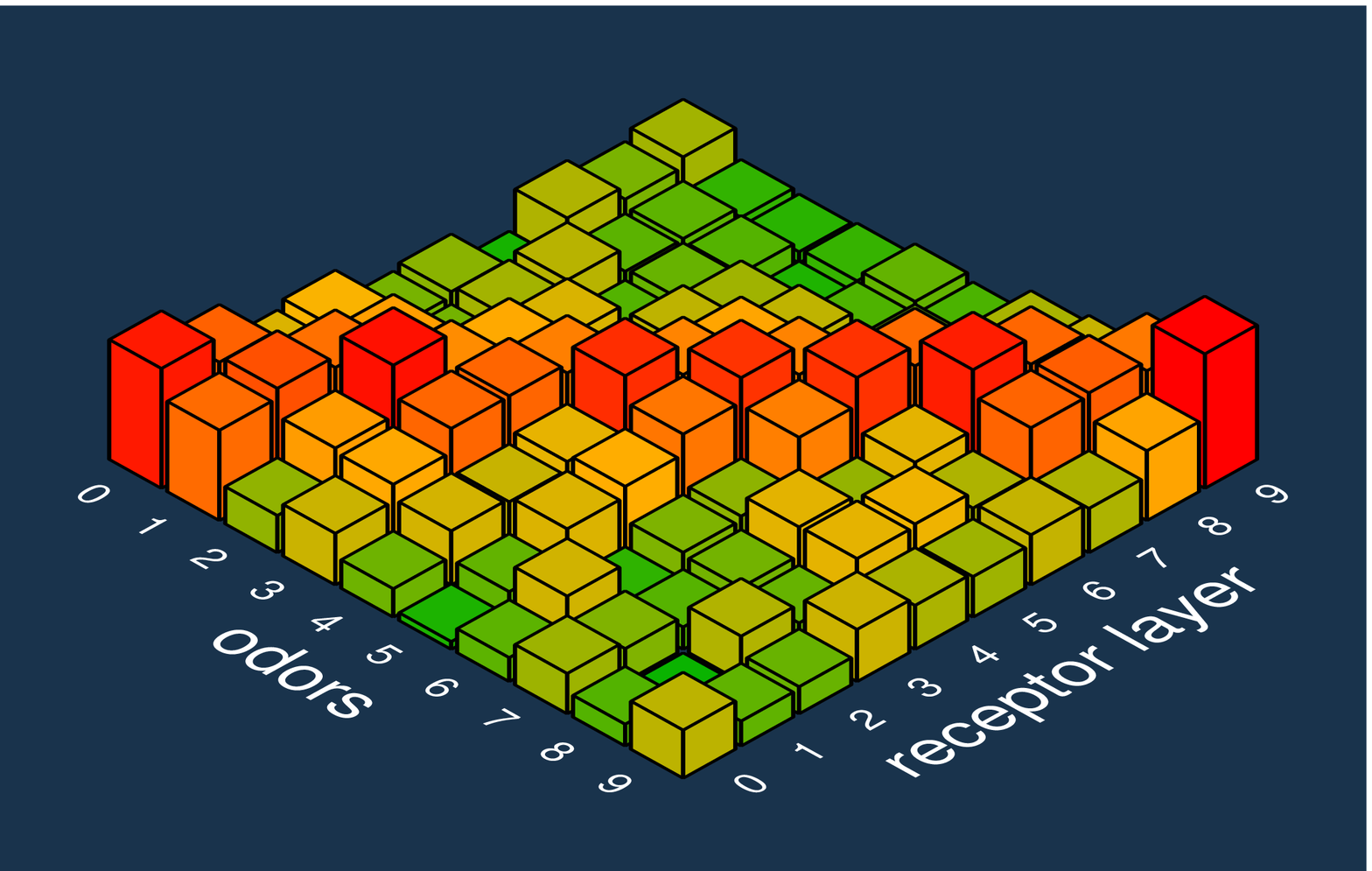}} \resizebox*{0.45\textwidth}{!}{\includegraphics{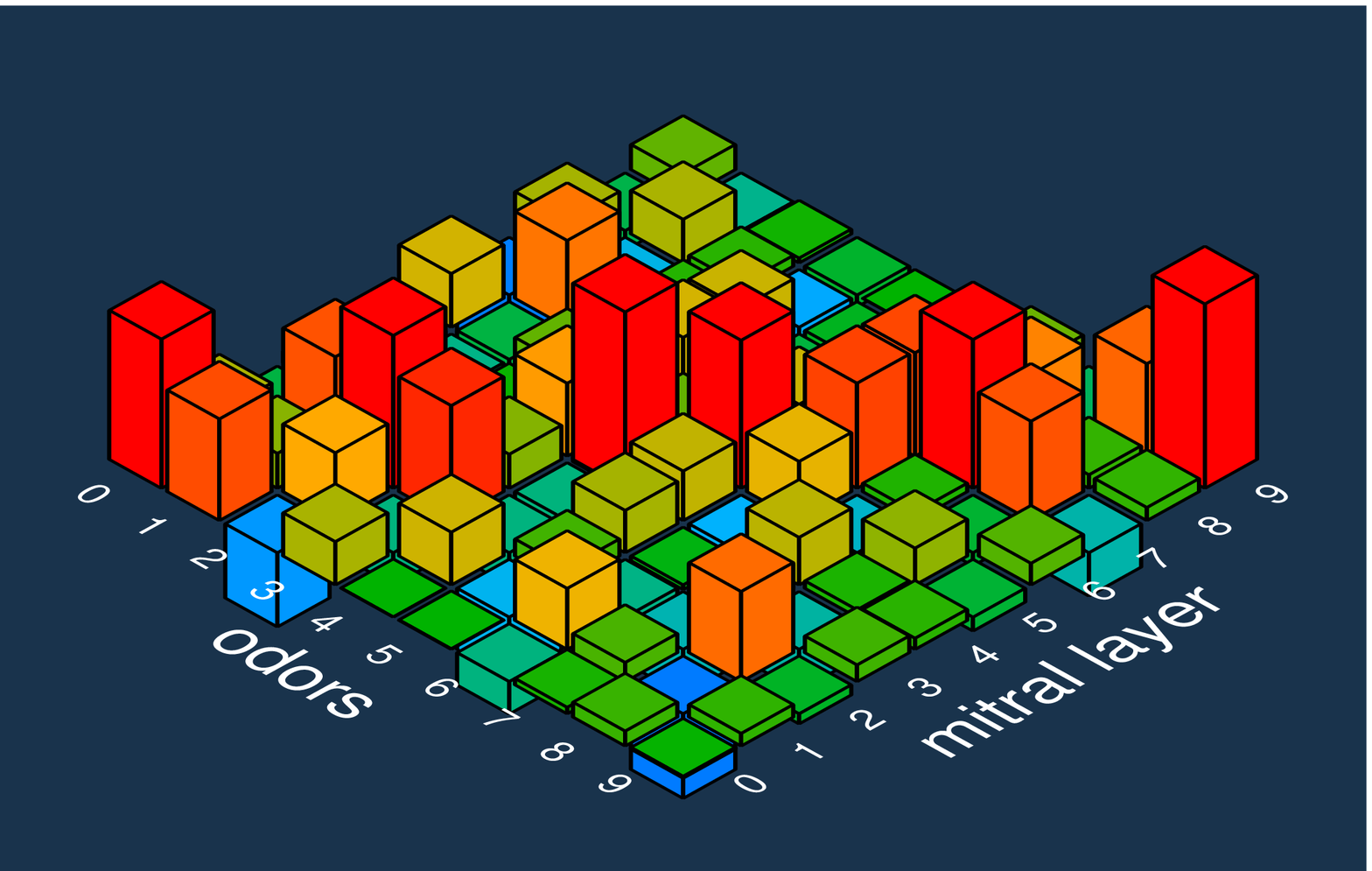}} \par}
\vspace{0.3cm}

\caption{{\footnotesize Representation of the smell matrix before and after application of the proposed algorithm. Left panel: an instance of odor ensemble with different degrees of mixing between receptors. Right panel: result of application of the algorithm; the mitral cell layer organizes to segregate topographically the outputs.}}\end{figure}

\begin{figure}[htbp]

\vspace{0.3cm}
{\centering \resizebox*{0.45\textwidth}{!}{\includegraphics{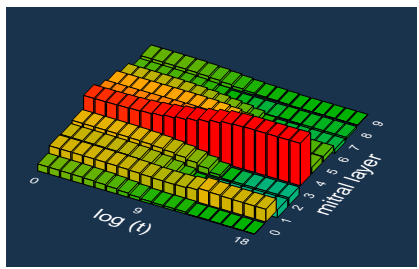}} \par}
\vspace{0.3cm}

\caption{{\footnotesize Evolution of the mitral cell layer response to one odor.}}\end{figure}

\begin{figure}[htbp]

\vspace{0.3cm}
{\centering \resizebox*{0.5\columnwidth}{!}{\rotatebox{-90}{\includegraphics{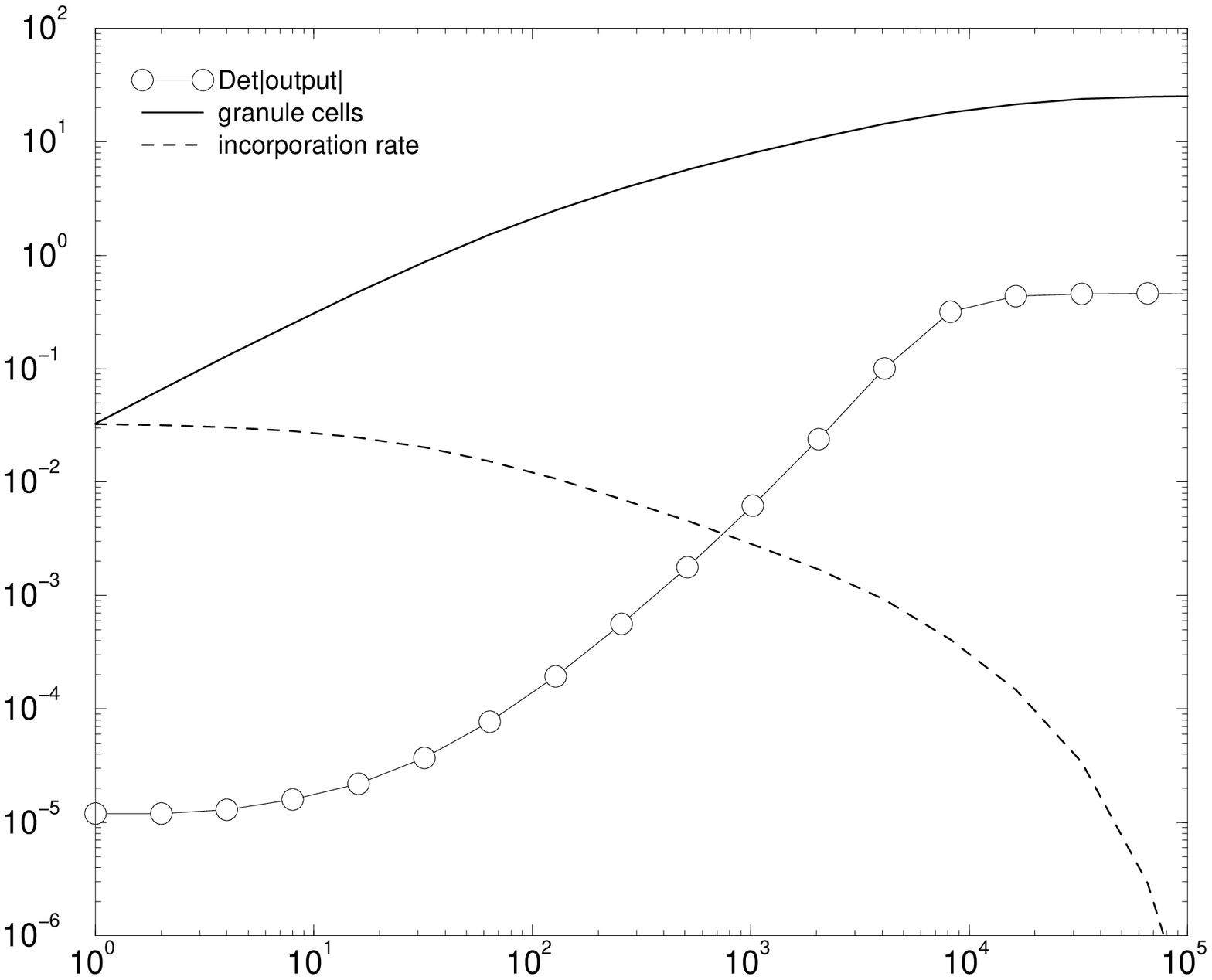}}} \par}
\vspace{0.3cm}

\caption{{\footnotesize Evolution of the network. Initially, the mitral cell layer matrix determinant equals the small odor matrix determinant. As the algorithm advances, the determinant (circle) as well as the total number of granular cells in the network (solid line) grow until a plateau is reached, after which the incorporation slows (dotted line, representing the derivative of granular cell incorporation) and finally approaches zero.}}\end{figure}

The other functional property of the algorithm is its adaptability to changes
in the input ensemble, i.e. a change in the set of odors the network is presented
with. This is a feature of paramount importance for early odor sensory processing,
given the intense seasonal changes produced, e.g., by blooming, migration and
foraging, and because the presence of a single odorant in high concentrations
is enough to upset the output of a system like the OB, due to the promiscuous
nature of the olfactory receptors. In our model, training is unsupervised, and
consists simply of exposing the system to the environment until a balance between
cell incorporation and death is established. Continued exposition to a constant
environment does not result in over-training. If the environment is changed,
the balance is upset and the system evolves until a new balance is achieved.
To see how our model copes with a changing environment, we simulated a periodic
renewal of the odor ensemble, on a time-scale similar to that of the convergence
of the orthogonalization process. The initial update rule was modified to include
a slow death process,
\begin{equation}
\label{eq5}
\Delta G_{ij}\sim \langle y_{i}y_{j}\rangle -\mu _{ij}
\end{equation}

\noindent where \( \mu _{ij} \) is a stochastic variable representing a constant
rate of \emph{unregulated} and \emph{non-specific} death, independent for each
granular unit, \( \mu _{ij}=\widehat{\mu }\: \: \forall ij \). The results
can be appreciated in Fig. 5, which depicts the evolution of the output determinant,
the total number of granular cells, and the number of granular cells connected
to one arbitrary pair of glomeruli. After the initial rise, the total number
of granular cells remains relatively constant during ensemble changes, whereas
any one particular location in the granular network fluctuates more tightly
with the changes. The orthogonalization process, although dependent on the identity
of the ensemble, is basically unaffected by the changes, as the determinant
is increased several orders of magnitude. Importantly, the non-specific death
term in the update equation, though small, is essential to assure adaptability.
Its absence leaves the network prone to preserve traces of previous ensembles,
which hinder the capability of orthogonalizing new ensembles (data not shown).

\begin{figure}[htbp]

\vspace{0.3cm}
{\centering \resizebox*{0.5\columnwidth}{!}{\rotatebox{-90}{\includegraphics{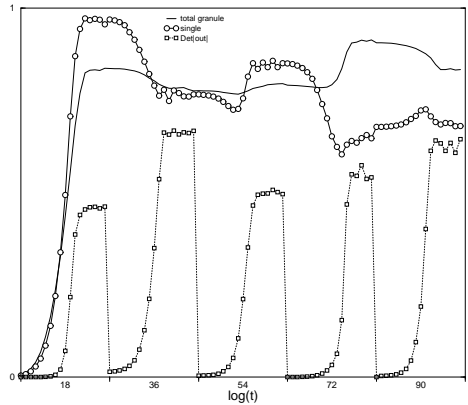}}} \par}
\vspace{0.3cm}

\caption{{\footnotesize Adaptation of neurogenesis to environmental changes. Each tick in the horizontal axis represents the exposure of the network to a new odor ensemble; the time axis is in a logarithmic scale which is reset after each new exposure. Symbols correspond to: ~solid line, total number of granular cells; circles, number of cells at one arbitrary position in the granular network; and squares, mitral cell layer matrix determinant.}}\end{figure}

\section*{Discusion}

We have modelled a simplified olfactory bulb circuit based on its known anatomy
and physiology in witch granule cells are constantly replaced. We have shown
that by having the young neurons to connect randomly and to survive in an activity
dependent basis, as experimental evidence suggests, neurogenesis is capable
of maximazing the discrimination of odors in a unsupervised manner. Moreover,
this discrimination task is robust to a changing environment by adding a small
random cell death rule.

Several physiological considerations are in order here. In the first place,
the implementation of the proposed survival rule implies only a \emph{local}
or somatic computation \cite{KOCH} (for instance, a biochemical mechanism involving
a high stoichiometry or cooperative reaction, such as the CamK-II/Ca\( ^{++} \)
pathway), and so requires much simpler molecular machinery than synapse-specific
modification mechanisms \cite{KANDEL}. One may think then that in evolutionary
terms, the selection of different strategies of plastic modification may depend
both on their computational possibilities and on ease of implementation, as
well as accessibility: neurogenesis was already available to build the structure
in the first place. 

The second consideration concerns the apparently hazaphardous distribution of
brain areas where adult neurogenesis has been documented. Although our model
was not intended to solve this puzzle, it suggests a possible answer. The replacement
of entire neurons is a rather violent event, in network terms, and rather unlike
smoother and more gradual rules like Hebbian changes or back-propagation. The
relationship between our algorithm and such gradual rules is similar to that
between Monte Carlo methods and gradient descent. In the former, the minimization
of an objective or cost function is done by randomly searching the configuration
space and thus avoiding trappings in local minima, whereas the latter performs
the minimization by downhill descent on the steepest directions of the function.
In these methods, the dimensionality and degree of smoothness of the function
to be optimized plays a substantial role as to which performs best. Thus we
hypothesize that the dimensionality of the input in competition with the amount
of computational resources dedicated to a task is what determines the existence
of adult neurogenesis. For instance, in the visual pathway there are several
layers of processing before reaching the cortex, whereas the OB, whose input
space has an intrinsic high dimensionality \cite{HOPFIELD}, projects directly
to the olfactory cortex. If synaptic plasticity is implemented by a diffusive
substance, a trophic factor for instance, then the diffusive process coupled
with the normal excursion of axonal and dendritic extensions can mislead the
guidance process. This will be particularly true if very distant locations in
the network must be connected, a very likely scenario in the case of olfaction,
where statistical correlations in the input do not possess a smooth topographic
projection, i.e. there is no simple sense of ``neighborhood''. According to
this hypothesis, the incorporation of new neurons can be correlated more precisely
with changes affecting global aspects of the input statistics.

We have presented the first mechanistic model in which adult neurogenesis provides
the scaffold underlying a computational task, namely the orthogonalization of
the neural responses to changing odor ensembles. The model predicts the differential
incorporation of neurons in particular regions of the OB upon olfactory environmental
changes, the necessity of a background of constant non-specific cell death,
and the existence of a somatic survival signal dependent on activity. It also
suggests that the structure of the input space determines to some extent the
existence of this mechanism in particular brain areas, thus providing a series
of predictions for future experimental manipulation.

\subsection*{Methods}

The update of the granular layer was computed from Eq. \ref{eq3} with \( \Delta G_{ij}=\gamma \langle y_{i}y_{j}\rangle  \)
where the average is computed over the cycle of presentations of the odors;
\( \gamma =5\times 10^{-3} \) in all simulations. The steady-state solution
is computed for each odor, and then normalized to \( |\vec{y}|=1 \). The orthogonalization
is quantified in Fig. 4 by computing the rank-1 determinant \( \Delta _{\cal O} \)
of the output matrix \( {\cal O}=\{{\bf y}^{k}\} \), where \( k \) ranges
over the ensemble of odors. In general, \( \Delta _{A} \) is the determinant
of the largest non-singular submatrix of \( A \), This can be computed for
any rectangular matrix from the singular value decomposition \( A=U\Lambda V \),
as \( \Delta =\prod _{\Lambda _{ii}\neq 0}\Lambda _{ii} \); in our simulations
has always dimension 10. The evolution of the total number of granular cells
in the network shows a monotonic increase, although the incorporation rate slows
dramatically after the stabilization of the orthogonalization is reached. For
the purpose of this demonstration, the rate of non-specific death was kept equal
to zero. For the simulation of Eq. \ref{eq5} depicted in Fig. 5 the random
death is defined as a stochastic variable which can take the values \( \widehat{\mu }=5\times 10^{-3} \)
with probability \( p=5\times 10^{-3} \) and \( \widehat{\mu }=0 \) with probability
\( 1-p \) in each iteration of the algorithm.

\subsubsection*{Acknowledgements}

The authors want to acknowledge fruitful discussions with T. Gardner, M. Sigman
and D. Chialvo, as well as comments on the manuscript by C. Scharff, L. Wilbrecht
and A.J. Hudspeth. Work supported in part by the Winston Foundation (GAC), Mathers
Foundation (MOM) and NICHD 32116 (AAB).


\begin{thebibliography}{}
\bibitem{COMBO}Rules in which granular cells connect to more than two neurons and whose survival
depends on the product of multiple inputs lead to combinatorial encoding strategies.
\end{thebibliography}
\end{document}